\documentclass[12pt]{article}
\usepackage{amsmath}
\usepackage{graphicx}
\usepackage{enumerate}
\usepackage{natbib}
\usepackage[title]{appendix}
\allowdisplaybreaks
\usepackage{url} 
\usepackage{tikz}
\usetikzlibrary{positioning, arrows.meta}
\usepackage{caption}

\newcommand{\blind}{1}
\newcommand{\cip}{\perp\!\!\!\perp}

\begin{document}

\def\spacingset#1{\renewcommand{\baselinestretch}%
{#1}\small\normalsize} \spacingset{1}



\def\spacingset#1{\renewcommand{\baselinestretch}%
{#1}\small\normalsize} \spacingset{1}

\if1\blind
{
  \title{\bf Chasing Shadows: How Implausible Assumptions Skew Our Understanding of Causal Estimands}
  \author{Stijn Vansteelandt and Kelly Van Lancker\\
    Department of Applied Mathematics, Computer Science and Statistics, \\ Ghent University
    }
    \date{}
  \maketitle
} \fi

\if0\blind
{
  \bigskip
  \bigskip
  \bigskip
  \begin{center}
    {\LARGE\bf  Chasing Shadows: How Implausible Assumptions Skew Our Understanding of Causal Estimands}
\end{center}
  \medskip
} \fi

\bigskip

\spacingset{1.5} 


The ICH E9 (R1) addendum on estimands, coupled with recent advancements in causal inference, has prompted a shift towards using model-free treatment effect estimands that are more closely aligned with the underlying scientific question. This represents a departure from traditional, model-dependent approaches where the statistical model often overshadows the inquiry itself. While this shift is a positive development, it has unintentionally led to the prioritization of an estimand's ability to perfectly answer the key scientific question over its practical learnability from data under plausible assumptions. We illustrate this by scrutinizing assumptions in the recent clinical trials literature on principal stratum estimands, demonstrating that some popular assumptions are not only implausible but often inevitably violated. We advocate for a more balanced approach to estimand formulation, one that carefully considers both the scientific relevance and the practical feasibility of estimation under realistic conditions.


\section{Introduction}

The ICH E9 (R1) addendum on estimands, along with advancements in causal inference, has prompted researchers to delve deeper into formulating scientifically meaningful questions. A shift has emerged towards the use of model-free estimands - measures of treatment effect that are well-defined without a strict reliance on a correctly specified statistical model. This makes it possible for estimands to be more closely tailored to the scientific inquiry, marking a departure from traditional, heavily model-based methods where the model often takes precedence over the actual scientific question \citep{vansteelandt2021statistical, kahan2024estimands}.

While we applaud this transformative trend, we urge for caution. Discussions surrounding the choice of estimand often prioritize which estimand provides the `ideal knowledge' needed to answer the scientific question, rather than critically evaluating whether and how the estimand can be effectively learned under assumptions that have a reasonable degree of plausibility.
The tripartite framework presented by \cite{akacha2017estimands} is a notable example, offering general recommendations on the choice of estimands based on their presumed relevance for various stakeholders, but ignoring that their estimation may demand implausible assumptions. 
Moreover, subsequent development of estimation strategies within this framework (e.g., \cite{qu2020general}) often proceeds with insufficient scrutiny of the underlying causal assumptions, the plausibility of which we demonstrate to be weak. This practice persists despite the long-standing tradition in causal inference to clearly state the assumptions - and assess their plausibility - needed to translate an estimand into a quantity that can be reliably inferred from observed data. 

We argue that the tension between theoretical clarity in acknowledging assumptions and the practical tendency to overlook their plausibility may be partly due to the complexity of the powerful counterfactual framework used for formalization \citep{hernan2010causal}. This is not a critique of the framework - which has driven major advances in causal inference - but rather a call to move beyond merely stating assumptions, towards carefully interpreting them in the context of the specific study at hand. Indeed, the mathematical intricacies of the counterfactual framework can obscure the potential unrealistic nature of certain causal assumptions. This poses a risk of formulating assumptions that appear standard and intuitive - examples of which will be given in subsequent sections - but are essentially guaranteed to be incorrect. Principal stratum estimands are especially - though not exclusively (see further) - vulnerable to this, because they are so difficult to learn from the observed data. To illustrate this point, we discuss several implausible assumptions commonly found in the (clinical trials) literature on principal stratification (see e.g., \cite{hayden2005estimator,qu2020general, bornkamp2021principal, lipkovich2022using, luo2022estimating}) and conclude by calling for a more cautious approach.

\section{Misguided ignorability}

\begin{figure}[h!]
\centering
\begin{tikzpicture}[node distance=2cm, >={Stealth[round]}, thick]

    \node (T) {T};
    \node (X) [above=of T] {X};
    \node (A) [right=of T] {A};
    \node (Z) [above=of T, right=of X, xshift=-0.75cm] {Z};
    \node (Y) [right=of A] {Y};
    \node (U) [draw,above=of Y, right=of Z, circle] {U};

    \draw[->] (X) -- (Y);
    \draw[->] (X) -- (Z);
    \draw[->] (X) -- (A);
    \draw[->] (T) -- (Z);
    \draw[->] (T) -- (A);
    \draw[->] (T) to[out=-40, in=-150] (Y);
    \draw[->] (Z) -- (A);
    \draw[->] (Z) -- (Y);
    \draw[->, red] (A) -- (Y);
    \draw[->] (U) -- (Z);
    \draw[->] (U) -- (Y);
    \draw[->] (X) to[out=40, in=140] (U);
    \draw[->] (T) -- (U);
    
\end{tikzpicture}
\caption{Causal diagram visualizing the causal relations between $X,T,Z,A$ and $Y$, allowing for an unmeasured confounder $U$ of $Z$ and $Y$.}\label{fig: DAG_A5}
\end{figure}
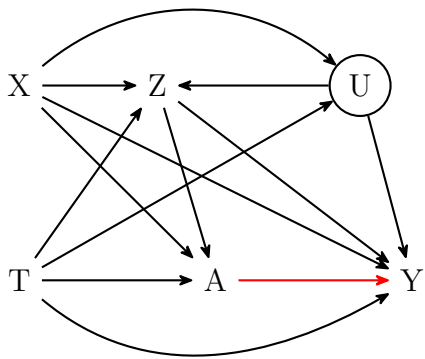

Facing complications due to treatment non-adherence, \cite{qu2020general}  propose estimation methods for evaluating the treatment effect for those who can adhere to one or both treatments. In their notation, $Y$ denotes the final outcome, $X$ represents a baseline covariate vector, $Z$ is a vector of intermediate post-baseline measurements, $T$ is the randomized treatment indicator ($T = 0$ for the control group and $T = 1$ for the experimental treatment group), and $A$ signifies the adherence status for the assigned treatment over the planned trial duration, where $A = 1$ implies that a patient completes the trial while adhering to the assigned treatment and observes the primary endpoint. Importantly, $Z$ plays the role of `pre-adherence' covariates: e.g., side effects such as injection
site reaction (AE-Inj) and prognostic factors such as HbA1c,
low density
lipoprotein cholesterol (LDL-C), triglyceride (TG), fasting
serum glucose (FBG) and alanine aminotransferase (ALT). These covariates may be affected by treatment and may in turn affect adherence.  This is graphically visualized in the (simplified) causal diagram of Figure \ref{fig: DAG_A5}. 

To formalize assumptions, we will use the notation $Y(t), A(t), Z(t)$ to denote the counterfactual or potential outcome, adherence status, and intermediate measurements under assignment to treatment level $t=0,1$. For any given patient, there exist two potential outcomes, $Y(1)$ and $Y(0)$, two potential adherence statuses, $A(1)$ and $A(0)$, and two potential intermediate measurements, $Z(1)$ and $Z(0)$. Typically, only one of $Y(1)$ or $Y(0)$ (similarly for $A(1)$ and $A(0)$, and $Z(1)$ and $Z(0)$) is observable, with the unobservable counterpart referred to as a counterfactual.

One key assumption made in some of the clinical trials literature on principal stratum estimands is that
\begin{equation}\label{eq:a5}
A(t) \cip \left\{Y(1), Y(0), Z(1-t)\right\}|X, Z(t) \quad \text{for} \quad t=0,1,\end{equation}
where $K\cip L|M$ for random variables $K,L$ and $M$ means that $K$ and $L$ are conditionally independent, given $M$. This assumption corresponds with Assumption A5 in \cite{qu2020general}; 
analogous assumptions are found in for instance \cite{qu2021implementation, luo2022estimating, qu2023assessing}, while \cite{bornkamp2021principal} and \cite{lipkovich2022using} make the similar assumption that $Y(t)\cip A(1-t)|X$. The latter assumption, $Y(t)\cip A(1-t)|X$, is often characterized as a `(weak) principal ignorability assumption' \citep{feller2017principal}, while the technical assumption in Equation (\ref{eq:a5}) has been referred to as an `ignorable adherence assumption' \citep{qu2020general}, in the sense that `adherence only depends on observed values $X$ and $Z(t)$'.
Since ignorability assumptions are routinely used in the causal inference and missing data literature, the label `ignorability' 
can misleadingly suggest that these assumptions are inherently plausible. We will argue below that this perceived plausibility of (\ref{eq:a5}) is misplaced. More critically, interpreting Equation (\ref{eq:a5}) as implying that `adherence depends only on observed values $X$ and $Z(t)$' is fundamentally incorrect: as we will demonstrate, Assumption (\ref{eq:a5}) can - and often will - be violated even when adherence is indeed solely influenced by observed values $X$ and $Z(t)$.

To see this, note that Assumption (\ref{eq:a5}) implies in particular that 
\begin{equation}\label{eq:a5bis}
A(t) \cip Y(t) |X, Z(t) \quad \text{for} \quad t=0,1,\end{equation}
which, by randomization (i.e., Assumption A4 in \cite{qu2020general}) is equivalent to assuming that
\begin{equation}\label{eq:a5obs}
A \cip Y |X, Z, T=t \quad \text{for} \quad t=0,1\end{equation}
(see Supplementary Materials, Section 1.1). 
 Assumption (\ref{eq:a5obs}) is much less obscure than (\ref{eq:a5}). It states that patients with poor adherence in treatment group $t$ have the same outcomes (in distribution) as patients with the same baseline covariates and (pre-adherence) intermediate covariates, but with good adherence in group $t$. This assumption is generally (guaranteed to be) violated, except (potentially) when the treatment is ineffective; \cite{feller2017principal} make a similar remark in their discussion of strong principal ignorability. It follows that also Assumption \eqref{eq:a5} (i.e., Assumption A5 in \cite{qu2020general}) is generally violated whenever adherence causally influences the outcome, which we may expect to be the case, even when adherence is solely influenced by observed values $X$ and $Z(t)$. 
Consider, for instance, the IMAGINE-3 Study analyzed by \cite{qu2020general}, a randomized Phase 3 trial comparing basal insulin peglispro with insulin glargine for type 1 diabetes. Patients who adhere closely to their insulin regimen are more likely to experience significant reductions in HbA1c levels by 52 weeks, while those with poor adherence might see less improvement.
We expect this to be the case, even when restricting to patients with the same baseline ($X$) and intermediate ($Z$) measurements of LDL-C, TG, FBG, ALT, and AE-Inj (as considered in \cite{qu2020general}).
This causal effect of adherence ($A$) on HbA1c ($Y$) is presented in the causal diagram in Figure \ref{fig: DAG_A5} as a red, direct arrow from $A$ to $Y$ (but is absent from Figure 1 in \cite{qu2020general}). Readers familiar with causal diagrams will note that Assumption \eqref{eq:a5bis} is indeed violated in the presence of a causal effect, no matter what collection of (pre-adherence) variables is being adjusted for (see Supplementary Materials, Section 2 
for further detail). Such effect of $A$ on $Y$ is likewise expected in most cases where $A$ refers to other intercurrent events than adherence, such as relapse, a biomarker exceeding a certain cut-off, \dots \citep{bornkamp2021principal}.

Assumption (\ref{eq:a5}) additionally implies
\[A(t) \cip Y(1-t) |X, Z(t) \quad \text{for} \quad t=0,1,\]
a condition more widely employed in the literature to identify certain principal stratum estimands (e.g., $E(Y(1)-Y(0)|A(t)=1)$ for $t=0,1$ and $E(Y(1)-Y(0)|A(0)=1, A(1)=1)$). However, also this assumption is unlikely to hold because $A(t)$ and $A(1-t)$ are typically associated due to factors not captured by the measured intermediate covariates (see Supplementary Materials, Section 1.2 
as well as the discussion in the next section; \cite{feller2017principal} and \cite{wang2023sensitivity} make a related remark in discussions on weak principal ignorability).
For instance, adherence in the IMAGINE-3 Study is likely influenced not only by the intermediate covariates HbA1c, LDL-C, TG, FBG, ALT, and AE-Inj, but also by personality traits and behavioral characteristics that are not reflected in these clinical measurements. Therefore, adjusting for these covariates alone will generally not suffice to achieve conditional independence between $A(t)$ and $A(1-t)$. Since moreover $A(1-t)$ is generally associated with $Y(1-t)$, conditional on $X$ and $Z(t)$ (as explained in the previous paragraphs), we may generally expect a (conditional) association between $A(t)$ and $Y(1-t)$. 

In summary, the mathematical complexities involved in formulating assumptions in terms of counterfactuals, as well as labels such as `ignorability', can sometimes lead to the misinterpretation that these assumptions are relatively weak (e.g., that they can be made to hold by adjusting for sufficiently many variables). This is for instance evident in the work of \cite{qu2020general}, who falsely claim that `since we treat the confounded measurements after intercurrent events as missing, this assumption (i.e., Assumption (\ref{eq:a5})) is equivalent to ignorable missingness or missing at random (MAR)'; this would be correct if adherence did not affect outcome, but if that were the case, one might as well ignore it. Similarly, \cite{hayden2005estimator} and \cite{qu2023assessing} incorrectly refer to $A(t) \cip \{Y(1-t), A(1-t)\}|X$ as the `explainable nonrandom noncompliance/survival assumption'' in \cite{robins1998correction}. Further, \cite{lipkovich2022using} mistakenly interpret $Y(t) \cip \left\{A(1), A(0)\right\}|X$ for $t=0,1$ as comparable to the assumption in propensity-based methods with observational data that `potential outcomes are independent of the non-randomly assigned treatment $T$ given covariates'. However, the parallel between Assumption \eqref{eq:a5} and MAR -- likewise for the parallels listed for the other assumptions -- cannot be drawn because the act of missingness typically does not influence the outcome, whereas adherence does. It is precisely this distinction between missingness assumptions and causal assumptions that necessitates the use of counterfactuals $Y(t,a)$ or $Y(a)$ \textit{indexed by the adherence status $a$}, representing the outcome that would be observed if (treatment were set to $t$ \textit{and}) adherence were set to $a$, as opposed to the use of counterfactuals $Y(t)$ \textit{indexed by the treatment status $t$}. In particular, in the causal inference literature, the assumption of ignorable adherence is formalized as
\[A(t) \cip Y(t,a)|X, Z(t) \quad \text{for} \quad t=0,1,\forall a,\]
or 
\[A \cip Y(a)|X, Z, T=t  \quad \text{for} \quad t=0,1,\forall a;\]
the `explainable nonrandom noncompliance/survival assumption' in \cite{robins1998correction} is an immediate generalization of this to multiple visit times.
Although this assumption resembles Assumption \eqref{eq:a5} (or a component of it), it is substantively different because $A(t)$ does not affect $Y(t,a)$ when adherence is fixed at $a$ (and likewise $A$ does not affect $Y(a)$), thereby now rightly allowing a parallel with MAR to be drawn.

\section{Misguided independence}

A second key assumption commonly made in recent principal stratification literature is
\begin{equation}\label{eq:a7}
Z(0) \cip Z(1) |X.\end{equation}
This assumption corresponds with Assumption A7 in \cite{qu2020general}, and is also found in other works such as \cite{qu2021implementation} and \cite{luo2022estimating}; the same assumption with $A$ in lieu of $Z$ is for instance found in \cite{hayden2005estimator}, \cite{lipkovich2022using} and \cite{qu2023assessing}.
We agree with \citet{lipkovich2022using} that `this assumption is particularly strong, as it essentially assumes that the cross-world random
effects associated with the same patient are conditionally independent given baseline covariates, which like any other
cross-world assumptions cannot be verified from the data, where each patient receives only one treatment'. \cite{qu2020general} agree that `it is generally difficult to evaluate this assumption', but do not comment on its plausibility. More recently, \cite{qu2023assessing} judge it to be `not unreasonable' that potential outcomes for alternative treatments in the same patient be conditionally independent, given the measured covariates, despite being naturally correlated. In contrast, we argue that use of this assumption is problematic as it is essentially certain to be violated for several reasons, which we will now explain.

First, reasoning under Pearl's nonparametric structural equation model, for Assumption \eqref{eq:a7} to be valid the covariates $X$ must capture all prognostic factors of $Z$.
However, the idea that post-baseline covariates could be perfectly predicted using baseline covariates is highly improbable, particularly given the introduction of a treatment to which different patients may respond differently. Furthermore, if the baseline covariates $X$ perfectly predicted the time-varying covariates $Z$ (aside from random noise), then adjusting for time-varying covariates would provide no additional value, undermining the core rationale behind \cite{qu2020general}'s approach of adjusting for post-baseline covariates.

Second, 
consider the scenario where the treatment has no effect, and, in particular, does not impact the intermediate measurements ($Z$). In this case, $Z(0)=Z(1)$, which would inherently violate Assumption \eqref{eq:a7}. This is a significant concern, as we should not accept an analysis that is guaranteed to be invalid under the null hypothesis of no treatment effect.

Third, to generalize beyond the null scenario, consider the following analogy. Suppose we have two repeated measurements, $Z_j,j=1,2$, representing the variable $Z$ measured twice per patient. These could be framed as counterfactuals $Z_j(1)$ for patients receiving the treatment and $Z_j(0)$ for patients under the control condition. Now, if we were to assume that these repeated measurements are conditionally independent, specifically that
\begin{equation}\label{eq:a7bis}
Z_1(t) \cip Z_2(t) |X,\end{equation}
for $t=0$ or $t=1$, this assumption would rarely, if ever, be tolerated in statistical analysis: it is nearly always biologically implausible, as supported by extensive evidence from repeated measures studies.
Even so, Assumption (\ref{eq:a7}) is arguably stronger than (\ref{eq:a7bis}). This is because $Z(0)$ and $Z(1)$ refer to measurements for the same individual \textit{at the same time}, albeit under different treatment conditions. In contrast, $Z_1(t)$ and $Z_2(t)$ refer to measurements taken at different times, possibly several months apart. Hence, assuming conditional independence between $Z(0)$ and $Z(1)$ is significantly more stringent and less justifiable than assuming it between repeated measures at different times under the same treatment, because a patient's condition, biology, environment and disease progression can be expected to influence the outcomes under both treatment conditions.
Assumption (\ref{eq:a7bis}) would generally be deemed unacceptable in a repeated measures analysis, despite violation of it generally not inducing bias in the estimation of treatment effects on $Z$. 
Given that Assumption (\ref{eq:a7}) is arguably much stronger, and that its violation does not merely affect standard errors, but primarily induces bias in estimators of principal stratum effects, it raises significant concerns to the point that analyses based on it should be considered unacceptable.

A fourth concern is that as more post-baseline covariates are being adjusted for (increasing the dimension of $Z$), Assumption (\ref{eq:a7}) becomes progressively stronger. This may lead investigators to limit adjustment to only a small set of post-baseline covariates. This stands in stark contrast to popular methods for adjusting for time-varying confounding \citep{hernan2010causal}, whose assumptions tend to weaken as more post-baseline covariates are included, thereby ‘rewarding’ investigators who collect and use rich information, with analysis results that have greater credibility \citep{vansteelandt2009discussion}.

In summary, while the central role of adjusting for common causes in causal inference might make assumptions like (\ref{eq:a7}) seem justifiable, it’s important to recognize a key distinction. Standard ignorability or exchangeability assumptions typically involve adjusting for all confounders affecting \textit{two different variables}, such as treatment and outcome. This may be  feasible when treatment decisions are based on a limited number of prognostic factors for the outcome.
However, Assumption (\ref{eq:a7}) demands something much more stringent: (a) comprehensive data on \textit{all} predictors of a \textit{single} variable,  $Z$, and (b) these predictors being available at baseline. This requirement is biologically implausible, as it is unlikely that all relevant predictors of $Z$ could be captured, and even if they could, it remains unlikely that no information past the start of the study additionally predicts $Z$.

\section{Misguided interpretation}

In our view, recommendations in the tripartite framework \citep{akacha2017estimands} on the choice of estimand  should be reconsidered, not only due to the aforementioned concerns about the plausibility of assumptions, but also because of oversimplification of the problem and misinterpretation of the principal stratum estimand, as we will explain next.

A common oversimplification is the reduction of patients into two categories: adherers and non-adherers. In reality, adherence is a complex, time-varying process where patients who adhere at one point may become non-adherent later. Any dichotomization of adherence risks unintended consequences. Beyond effect dilution, one such consequence is the challenge of interpreting the principal stratum of adherers in time-to-event studies \citep{mattei2024assessing,liu2024principal}, which make up a significant portion of clinical trials. Patients who die early are more likely to belong to the principal stratum of adherers, as the likelihood of non-adherence increases over time. This raises doubts about patients’ ability to self-classify as adherers, as doing so would require not only predicting their future adherence but also estimating their remaining lifespan.

Another oversimplification involves the use of natural language. For instance, \cite{akacha2017estimands} suggest that `any meaningful estimand should start with, `if this treatment is taken as labeled ....' However, this phrasing refers to a hypothetical estimand, not a principal stratum estimand. For the latter, the appropriate phrasing would be: `if you belong to the subgroup of patients who would adhere in a blinded randomized clinical trial, and you take the treatment as labeled.' This more subtle phrasing acknowledges that patients who would adhere in a clinical trial may not do so in the real world. Oversimplifying the language risks giving the impression that the principal stratum estimand is of broad interest when, in reality, it may not be.
Similarly, providing patients with the answer to the question posed in Section 2.1 of \cite{akacha2017estimands} - `If I can tolerate the treatment and continue to take it (at any dose), what efficacy response and side effects am I likely to have over the prescribed or recommended treatment duration?' oversimplifies the interpretation of a principal stratum estimand, and may moreover not be as relevant as suggested. Knowing the answer to this question could influence patient behavior in ways that affect adherence, potentially making the principal stratum of adherers in the clinical trial significantly different from that in the real world. 

\section{Conclusions}

\subsection{Key points}

The conclusions drawn by 
\cite{mealli2012refreshing}, \cite{qu2020general, qu2021implementation} and  \cite{luo2022estimating} emphasizes the broad applicability of principal stratum estimands in various randomized controlled trials, including questions beyond the primary endpoint in certain trials \citep{bornkamp2021principal}. While \cite{qu2020general}, among others, defend the greater complexity in inferring these by noting that getting a good answer to the right question is worth it, 
our standpoint introduces a note of caution. We contest this assertion in light of the absence of plausible assumptions that enable one to learn such effects from data. When selecting a suitable causal estimand, striking a balance becomes imperative between the right question and the feasibility to answer it under realistic assumptions. This balance is easier to achieve when one acknowledges the nuanced nature of the situation, which decision makers at regulatory agencies are well aware of \citep{kuemmel2020consideration}: there is rarely a single estimand that fully addresses the scientific question. We hereby align with the position of \cite{pearl2011principal} that `when comparing multiple estimands of similar value, identifiability becomes a key criterion for selecting a preferred estimand. If the assumptions required to identify estimand 1 are weaker or more realistic than those needed to identify estimand 2, this should be a crucial factor in deciding which estimand to focus on for inference.'

We have supported our concerns by scrutinizing assumptions in the recent clinical trials literature on principal stratum estimands, demonstrating that some popular assumptions are not only implausible but often inevitably violated. Concerns have likewise been voiced by scientific experts on principal stratification \citep{feller2017principal,wang2023sensitivity}, but in our experience are not given due attention. For instance, \cite{qu2023assessing} sought to evaluate the assumptions discussed in this article using a $2 \times 2$ cross-over study, where potential outcomes under both treatments were observable, albeit at different times.  Based on their results, they conclude that $A(t) \cip Y(1-t) |X$ for $t=0,1$ and Assumption \eqref{eq:a7} (w.r.t. adherence $A$ instead of $Z$) conditional on $X$ (as well as unconditionally) are quite reasonable for practical data analyses. While we welcome attempts to study the plausibility of assumptions, we disagree with their conclusions for several reasons, besides those listed in previous sections: (1) their evaluation hinges on the unrealistic assumption of no carry-over or period effects, \textit{even at the individual level}, (2) the small sample sizes in their study raise concerns about the statistical power necessary to rigorously test these assumptions, (3) the methodology they employed contradicts the fundamental principle that one should never accept the null hypothesis, and (4) even if the conclusion were true, no general recommendations for other studies should be based on it. 

Our focus on principal stratum estimands has been motivated by the common misinterpretations and limited plausibility of the assumptions mentioned earlier in the clinical trials literature (see also \cite{dukes2021identification}). Other causal analyses aimed at estimands that are likewise difficult to learn from the observed data, may be similarly vulnerable to these issues. Indeed, the introduction of a powerful counterfactual framework for identifying causal estimands has spurred an industry of causal inference developments under mathematical convenience assumptions that are challenging to interpret or explain. Consequently, these assumptions are seldom scrutinized in illustrative data analyses. We therefore advocate for causal analyses that go beyond merely stating causal assumptions. It is essential to also interpret and critically evaluate these assumptions in the context of substantive applications, and otherwise to be explicit about the possible dangers of relying on such assumptions.

\subsection{Estimand relevance}

If no plausible assumptions can be identified, it is often a sign that the chosen estimand is too disconnected from the real-world context to be directly valuable for decision-making. For instance, understanding the efficacy of a treatment for patients capable of adhering to the new regimen until the end of the study becomes limited in utility if we don't know or fully understand other critical factors.
These uncertainties include identifying these adherent patients prior to the start of treatment (at the time when decisions about treatment assignment need to be made), discerning potential harms in other patient groups (i.e., in other principal strata under study) \citep{mealli2012refreshing}, and acknowledging the variability in the stratum of patients capable of adherence across studies due to differences in study duration and mortality risks \citep{dawid2012imagine}.

This concern extends beyond principal stratum estimands to hypothetical estimands, which express treatment effects under the assumption that all patients adhere until the study's end. Ambitious attempts to infer such estimands become challenging in scenarios where patients are prematurely withdrawn from treatment following adverse events. This results in violations of positivity, as patients who discontinued due to adverse events usually lack comparable counterparts who remained adherent. This is particularly true when adherence decisions follow a (near-)deterministic rule outlined in the study protocol, though it can also occur in other contexts.
In such instances, a careful consideration of estimands that extrapolate closer to the available data \citep{young2014identification,michiels2021novel,rudolph2022effects} or alternative analyses invoking different causal assumptions \citep{michiels2024adjusting} becomes necessary.

The practical value of basing treatment recommendations on the treatment effect for a principal stratum has been questioned by many \citep{dawid2012imagine,joffe2011principal, sjolander2011reaction, vanderweele2011principal, 
prentice2011invited, 
pearl2011principal, stensrud2022translating}, particularly given that we can never identify which patients would adhere regardless of their treatment assignment. While it is sometimes recommended to use baseline covariates to predict membership in this principal stratum \citep{roy2008principal, ich2019, lipkovich2022using}, such predictions will always be imperfect. Targeting the treatment effect in the group of patients - identifiable before treatment begins - whose estimated probability of belonging to that principal stratum exceeds a chosen threshold, likewise has limited utility since clinicians will unlikely ever have access to such probability estimates.

Alternatively, some researchers have found appeal in treatment effects for `identifiable' unions of principal strata \citep{qu2020general}, such as patients who would adhere if assigned to treatment (disregarding whether they would also adhere on control). This has the drawback that it does not entirely balance adherence across both arms of the trial, but the advantage that these patients can be identified at the study's conclusion. In spite of this, the practical relevance of the resulting principal stratum effects remains questionable as it continues to be uncertain in a prospective setting - where treatment decisions must be made for new patients - whether a particular patient belongs to the stratum of patients who would adhere if assigned to treatment. This makes it impossible to administer treatment in a way that targets only those patients, which is especially a concern in case treatment is not beneficial for patients outside this stratum. 

\subsection{Recommendations}

For the pharmaceutical industry to report estimands that are truly relevant to real-world practice, they must carefully consider how the treatment will be recommended for use in practical settings. If perfect adherence is recommended, regardless of the presence of side effects, it is well justified to report the hypothetical estimand corresponding to perfect adherence, even if it is challenging to estimate from the observed data. In the latter case, it may be beneficial to supplement this with a policy estimand or, if adherence levels in the trial do not align with those in real-world practice, with an evaluation of how well the treatment would perform if non-adherence levels were shifted to match those observed or anticipated in the target population.
If rescue treatment is provided to patients with poor prognosis, then treatment recommendations should include the option to switch patients to rescue treatment under specific conditions, accompanied by an evaluation of the effect of the resulting dynamic treatment strategy; see \cite{stensrud2022translating} for related recommendations. Similarly, if treatment recommendations suggest discontinuing treatment for nonresponders, the treatment effect for the principal stratum of responders becomes less useful. Instead, the focus should be on evaluating the effect of a dynamic treatment regime in which treatment is initially administered to all patients, and then discontinued for those who are judged to respond insufficiently. Such policy evaluations are directly informative about study designs that for instance re-randomize nonresponders. 
While such evaluations 
might traditionally be viewed as beyond the scope of work by the pharmaceutical industry, we argue that when treatments are brought to market, they should be accompanied by clear guidelines not only regarding dose, formulation, and similar factors but also on how to manage key intercurrent events. The reported treatment effects must then be consistent with these guidelines.

\section*{Acknowledgements}

We thank Oliver Dukes, Johan Steen and the participants of the Ghent University causal research working group for helpful discussions, and
an associate editor for valuable feedback. The authors report there are no competing interests to declare.

\bibliographystyle{chicago}
\bibliography{NISS-discussion}

\end{document}


\def\spacingset#1{\renewcommand{\baselinestretch}%
{#1}\small\normalsize} \spacingset{1}



\def\spacingset#1{\renewcommand{\baselinestretch}%
{#1}\small\normalsize} \spacingset{1}

\if1\blind
{
  \title{\bf Supplementary Material for `Chasing Shadows: How Implausible Assumptions Skew Our Understanding of Causal Estimands'}
  \author{Stijn Vansteelandt and Kelly Van Lancker\\
    Department of Applied Mathematics, Computer Science and Statistics, \\ Ghent University
    }
    \date{}
  \maketitle
} \fi

\if0\blind
{
  \bigskip
  \bigskip
  \bigskip
  \begin{center}
    {\LARGE\bf  Chasing Shadows: How Implausible Assumptions Skew Our Understanding of Causal Estimands}
\end{center}
  \medskip
} \fi

\bigskip

\spacingset{1.5} 

\section{Proofs}
\subsection{Ignorable adherence assumption: one world}\label{app:ignAdh}
To prove that \[A(t) \cip Y(t) |X, Z(t) \quad \text{for} \quad t=0,1,\] implies \[A \cip Y |X, Z, T=t \quad \text{for} \quad t=0,1.\] by randomization, we use graphoid axioms \citep{dawid1979conditional, pearl1988probabilistic}.

First, due to randomization, we know that $T\cip \left(Y(t), A(t), Z(t)\right) | X$ for $t=0,1$. The weak union law then implies $T\cip Y(t) | A(t), Z(t), X$ for $t=0,1$.  Next, by combining $A(t) \cip Y(t) |X, Z(t)$ with $T\cip Y(t) | A(t), Z(t), X$, the contraction law yields $Y(t) \cip \left(T, A(t) \right)|X, Z(t)$.  The weak union law further provides $Y(t) \cip A(t) |X, Z(t), T$. Setting $T=t$, we obtain $Y(t) \cip A(t) |X, Z(t), T=t$, which by consistency is equivalent to $Y \cip A |X, Z, T=t$. 

\subsection{Ignorable adherence assumption: cross-world}\label{app:ignAdh2}
To explain the restrictiveness of 
\[A(t) \cip Y(1-t) |X, Z(t) \quad \text{for} \quad t=0,1,\]
we express the associated non-parametric structural equation model (NPSEM):
\begin{align*}
   Z(t) &= f_Z(t, X, \epsilon_Z)\\
   A(t) &= f_A(t, X, Z(t), \epsilon_A)\\
   Y(t) &= f_Y(t, X, Z(t), A(t), \epsilon_Y)\\
   A(1-t) &= f_A(1-t, X, Z(1-t), \epsilon_A)\\
   Y(1-t) &= f_Y(1-t, X, Z(1-t), A(1-t), \epsilon_Y).
\end{align*}
Then, in general, $A(t)\cdp Y(1-t) |X, Z(t)$ as $Y(1-t)$ is a function of $A(1-t)$ which shares an error term, $\epsilon_A$, with $A(t)$. So, if $X$ and $Z(t)$ are not sufficient to cancel this noise, $A(t)$ and $A(1-t)$ will be associated and so will $A(t)$ and $Y(1-t)$, violating the assumption. This dependence will not be present if either $Y(1-t)$ is not a function of $A(1-t)$ or if there is no error term $\epsilon_A$. The latter would mean that $A(t) \cip A(1-t) |X, Z(t)$ for $t=0,1$.

\subsection{Proof of \cite{qu2020general}}
In this appendix we add detail to a proof by \cite{qu2020general} to show how Assumptions (1) and (4) from the main text 
are used for the identification of $E\{Y(1)-Y(0)|A(1)=1\}$ in their Appendix `A.2. Population That Can Adhere to the Experimental Treatment'. We hereby also focus on $E\{Y(0)|A(1)=1\}$. 

By the law of total expectation, $E\{Y(0)|A(1)=1\}$ can be rewritten as 
\[\frac{E\{A(1)Y(0)\}}{P(A(1)=1)}=
\frac{E\{A(1)Y(0)\}}{E\{A(1)\}}.\]
Under Assumption (1) from the main text, 
more specifically $A(1) \cip Y(0) |X, Z(1)$, this equals
\[\frac{E[\{P(T=1|X)\}^{-1}I(T=1)A(1)E\{Y(0)|X, Z(1)\}]}{E[\{P(T=1|X)\}^{-1}A(1)T]}.\]
Assuming $Y(t) \cip Z(1-t) |X, Z(t)$ and consistency, this is equivalent to
\[\frac{E[\{P(T=1|X)\}^{-1}I(T=1)A(1)E[E\{Y(0)|X, Z(0)\}|X, Z(1)]]}{E[\{P(T=1|X)\}^{-1}AT]}.\]
Under consistency and Assumption (4) from the main text, 
this corresponds to
\[\frac{E[\{P(T=1|X)\}^{-1}I(T=1, A=1)E[E\{Y(0)|X, Z(0)\}|X]]}{E[\{P(T=1|X)\}^{-1}AT]}.\]
Similar (identification) assumptions are used for the inverse probablity weighting estimator.

\section{Single World Intervention Graph}\label{app:addFig}

\begin{figure}[h!]
\centering
\begin{tikzpicture}[node distance=2cm, >={Stealth[round]}, thick]

    \node (T) {T\hspace{0.25cm} t};
    \node (X) [above=of T] {X};
    \node (A) [right=of T] {A(t)};
    \node (Z) [above=of T, right=of X, xshift=-0.75cm] {Z(t)};
    \node (Y) [right=of A] {Y(t)};

    \draw[->] (X) -- (Y);
    \draw[->] (X) -- (Z);
    \draw[->] (X) -- (A);
    \draw[->] (T) -- (Z);
    \draw[->] (T) -- (A);
    \draw[->] (T) to[out=-40, in=-150] (Y);
    \draw[->] (Z) -- (A);
    \draw[->] (Z) -- (Y);
    \draw[->, red] (A) -- (Y);
    
\end{tikzpicture}
\caption{SWIG to explain why $A(t) \cip Y(t) |X, Z(t)$, Assumption A5 in \cite{qu2020general}, is likely to be violated. Reasoning does not change when adding an unmeasured confounder of $Z$ and $Y$.}\label{fig: SWIG_A5}
\end{figure}
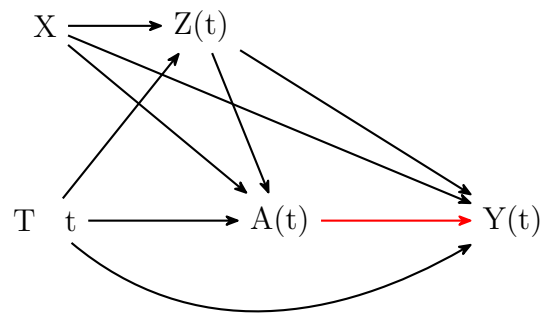

The Single World Intervention Graph (SWIG) in Figure  \ref{fig: SWIG_A5} corresponds with the causal DAG of Figure 1 in the main text. 
It can be used to directly verify, using d-separation, that $A(t) \cip Y(t) |X, Z(t)$ is generally violated.

\bibliographystyle{chicago}
\bibliography{NISS-discussion}